\documentstyle[manuscript,aps,prl]{revtex}

\begin{document}
\draft
%\twocolumn
\input{epsf}

\title{Ellipsometric measurements by use of photon pairs generated by spontaneous parametric down-conversion}

\author{Ayman F. Abouraddy, Kimani C. Toussaint, Jr., Alexander V. Sergienko,
Bahaa E. A. Saleh, and Malvin C. Teich }
\address{Quantum Imaging Laboratory, Departments of Electrical $\&$ Computer Engineering and Physics, Boston University, Boston, MA $02215-2421$}
%\email{raddy@bu.edu}
\maketitle

\begin{abstract}We present a novel interferometric technique for
performing ellipsometric measurements. This technique relies on
the use of a non-classical optical source, namely,
polarization-entangled twin photons generated by spontaneous
parametric down-conversion from a nonlinear crystal, in
conjunction with a coincidence-detection scheme. Ellipsometric
measurements acquired with this scheme are absolute; i.e., they do
not require source and detector calibration.
\end{abstract}

\noindent Ellipsometry \cite{PD,RA,HT,AR,AW,MM} is a
well-established metrological technique that is used, particularly
in the semiconductor industry, to determine the thickness and
optical constants of thin films. Because of the high accuracy
required in measuring these parameters, an \textit{ideal}
ellipsometric measurement would require absolute calibration of
both the source and the detector. As this is not attainable in
practical settings, ellipsometry makes use of a myriad of
experimental techniques developed to circumvent the imperfections
of the devices involved. The most common techniques are
\textit{null} and \textit{interferometric} ellipsometry.

In this paper we demonstrate how to satisfy the aforementioned
requirements by using a non-classical source of light: a
two-photon polarization-entangled state generated by type-II
spontaneous parametric down-conversion (SPDC)\cite{PK,DK1}. This
source has been used in the emerging field of 'quantum
metrology'\cite{DB}. We show that by utilizing the quantum
correlations exhibited by the photon pairs in such a state, one
may obtain absolute ellipsometric data from a sample. This is done
in a simple setting with a minimal number of optical components.

In the traditional (classical) null ellipsometer the sample is
illuminated with a beam of light that can be prepared in any state
of polarization. The reflected light, which is generally
elliptically polarized, is then analyzed. The polarization of the
incident beam is adjusted to compensate for the change in the
relative amplitude and phase between the two eigenpolarizations
introduced by the sample, such that the resultant beam is linearly
polarized. If it is passed through an orthogonal linear polarizer,
this linearly polarized beam will yield a null (zero) intensity at
an optical detector. The null ellipsometer does not require a
calibrated detector because it does not measure intensity, but
instead records a null. The principal drawback of null measurement
techniques is the need for a reference against which to calibrate
the null, for example, to find its initial location (the
rotational axis of reference at which an initial null is obtained)
and compare it with the subsequent location when the sample is
inserted into the device. The accuracy and reliability of all
measurements crucially depend on the accuracy and reliability of
our knowledge of the reference sample.

A traditional (classical) alternative is to employ an
interferometric configuration in which the light from the source
follows more than one path to the detector. The sample is placed
in one of those paths. The overall throughput of the system (both
source and detector parameters) is estimated by insertion of the
sample. This technique does not require a reference sample, but
relies on knowledge of the detector's efficiency. Interferometric
configurations employ more optical components (such as beam
splitters) that require accurate characterization and in general
increase instrumental errors.

The proposed entangled-photon quantum ellipsometer is illustrated
in Fig. 1. An intense laser (pump) beam illuminates a birefringent
nonlinear optical crystal (NLC). Quantum mechanics predicts that
some of the pump photons disintegrate into pairs, known
traditionally as signal and idler, which conserve energy
(frequency-matching) and momentum (phase-matching)\cite{DK1,BES}.

For our purposes, we choose the SPDC to be in a configuration
known as 'type-II non-collinear'. 'Type-II' refers to the fact
that the signal and idler photons have orthogonal polarizations
(ordinary and extraordinary) to satisfy the phase-matching
conditions; the term 'non-collinear' indicates that the signal and
the idler photons are emitted in two different directions. Because
of the birefringence of the NLC that is used to generate SPDC, the
signal and idler photons emerge from the NLC with a relative time
delay that one compensates for by placing an appropriate
birefringent material of suitable thickness in the signal and/or
idler path \cite {PK}. Specifically, in type-II SPDC from a
negative uniaxial NLC [such as beta-barium borate] the signal
photon (extraordinary polarized) emerges from the NLC before the
idler photon (ordinary polarized). One compensates for this by
placing a quartz plate, for example, of suitable thickness in one
of the arms of the setup. The exit surface of the NLC typically
has an antireflection coating to reduce reflection losses of the
two polarizations.

The signal and idler photons are emitted in a
polarization-entangled state described by \cite{PK,DK1}
\begin{equation}\label{entangled-state}
|\Psi\rangle=\frac{1}{\sqrt{2}}(|HV\rangle+|VH\rangle),
\end{equation}
where H and V represent horizontal and vertical polarizations,
respectively. It is understood that the first polarization in any
ket is that of the signal photon and the second is that of the
idler. Although the two-photon entangled state is a pure quantum
state, the signal and idler photons considered separately are each
unpolarized \cite{UF,AA1}. As shown in Fig. 1, the signal beam
encounters a linear polarization analyzer (A$_{1}$), followed by a
single-photon photodetector (D$_{1}$). The idler beam reflects off
the sample of interest before it encounters a linear polarizer
(A$_{2}$) followed by single-photon photodetector (D$_{2}$). The
sample is characterized by the parameters $\psi$ and $\Delta$:
$\psi$ is the ratio of the magnitudes of the sample reflection
coefficients, $R_{H}$ and $R_{V}$ for the p- and s-polarized
waves, respectively; $\Delta$ is the phase shift between them. The
detectors are part of a circuit that records the coincidence rate
of photon pairs.

The non-classical source and the optical arrangement shown in Fig.
1 exhibit two features that circumvent the two problems noted
above, i.e., calibration of the source and of the detector. The
first characteristic of our proposed quantum ellipsometer is that
the source is a twin-photon source; i.e., we are guaranteed on
detection of one photon in one of the arms of the setup that its
twin is in the other. The detection of one photon may be used to
gate the arrival of its twin in the other arm and thus we are
effectively provided with a calibrated optical source because the
efficiency of the gating detector is immaterial. The second
characteristic is the polarization entanglement of the source.
Polarization entanglement acts as an interferometer in our
approach, thereby alleviating the need for calibrating the second
detector in our coincidence scheme.

The coincidence rate, $N_c$, recorded by D$_{1}$ and D$_{2}$ is
proportional to the fourth-order coherence function, \cite{RG,BS},
and is given by
\begin{eqnarray}\label{coincidence-rate}
&N_{c}&=C|\beta\,e^{j\Delta}\cos\theta_{1}\sin\theta_{2}+\sin\theta_{1}\cos\theta_{2}|^{2}\nonumber\\&&
\propto\beta^{2}\cos^{2}\theta_{1}\sin^{2}\theta_{2}+\sin^{2}\theta_{1}\cos^{2}\theta_{2}+2\beta\cos\Delta\cos\theta_{1}\sin\theta_{1}\cos\theta_{2}\sin\theta_{2}.
\end{eqnarray}
Here $C$ is a constant that includes the quantum efficiency of the
detectors and the various parameters of the experimental
arrangement, $\beta=\sqrt{\tan\psi}$, and $\theta_j$ $(j=1,2)$ is
the angle of the analyzer with respect to H. If the sample is
replaced by a \textit{perfect} mirror, the coincidence rate is a
sinusoidal pattern of 100$\%$ visibility. In practice, by
judicious control of the apertures placed in the down-converted
beams, visibilities close to 100$\%$ can be obtained.

One may use relation (\ref{coincidence-rate}) to extract
ellipsometric data by fixing one of the analyzers and rotating the
other. It is advantageous to fix analyzer A$_{2}$ in the sample
arm and rotate A$_{1}$. One may choose $\theta_{2}=45^{\circ}$,
for example, whereupon
\begin{equation}\label{coinc-rate-45}
N_{c}=\frac{C}{2}|\beta\,e^{j\Delta}\cos\theta_{1}+\sin\theta_{1}|^{2}.
\end{equation}
Three angles of A$_{1}$ are sufficient for estimating the three
parameters $C$, $\psi$, and $\Delta$ (an obvious choice would be
$\theta_{1}=0^{\circ}, 45^{\circ}, 90^{\circ}$). It is sometimes
advantageous to choose a different value for $\theta_{2}$ to
equalize the two terms in the first line of Eq.\
(\ref{coincidence-rate}), particularly if $\beta\gg1$ or
$\beta\ll1$.

An important feature of this interferometer is that it is not
sensitive to an overall mismatch in the length of the two arms of
the setup. In this case one can show that the coincidence rate is
identical to that given in Eq.\ (\ref{coincidence-rate})
regardless of the mismatch.

Another advantage of this setup over its idealized null
ellipsometric counterpart is that the two arms of the ellipsometer
are separate and the light beams traverse them independently in
different directions. This configuration allows various
instrumentation errors of the classical setup to be circumvented.
The advantage of placing all optics in the idler channel arises
from the fact that optical alignment is easier to achieve in a
transmission, rather than in a reflection, configuration. The
system is less prone to beam-deviation error than is its
counterpart \cite{JZ}. In our case no optical components are
placed between the source (the NLC) and the sample; any desired
polarization manipulation may be performed in the other arm of the
entangled-photon ellipsometer.

A significant drawback of classical ellipsometry is the difficulty
of fully controlling the polarization of the incoming light. A
linear polarizer is usually employed at the input of the
ellipsometer, but the finite extinction coefficient of this
polarizer causes errors in the estimated parameters \cite{RA}. In
the quantum ellipsometer the polarization of the incoming light is
dictated by the phase-matching conditions of the nonlinear
interaction in the NLC. The polarizations defined by the
orientation of the optical axis of the NLC act as the input
polarization in classical ellipsometry. The NLC is aligned for
type-II SPDC, so only one polarization component of the pump
generates SPDC, whereas the orthogonal (undesired) component of
the pump does not (because it does not satisfy the phase-matching
conditions). The advantage is therefore that the down-conversion
process ensures the stability of polarization along a particular
direction.

There are also important practical advantages in employing quantum
ellipsometry.  All optics (in this setting only analyzer A$1$) is
placed in a path that does not include the sample. Because A$2$ is
the only optic in the sample arm, and as it follows the sample,
one can change the angle of incidence to the sample easily and
repeatedly.

An illuminating way to represent the action of the
entangled-photon quantum ellipsometer is provided by the redrawing
of Fig. 1 in the unfolded configuration shown in Fig. 2. Using the
advanced wave interpretation, which was suggested by Klyshko in
the context of two-photon imaging\cite{DK2}, one may obtain the
coincidence rate for photons at D$_{1}$ and D$_{2}$ by tracing
light waves originating from D$_{1}$ to the NLC and then onto
D$_{2}$ on reflection from the sample. With this interpretation,
the configuration in Fig. 2 becomes geometrically similar to that
of the classical ellipsometer. Although none of the optical
components usually associated with interferometers (beam splitters
and wave plates) is present in this scheme, interferometry is
still performed through the entanglement of the source and
coincidence measurements.

In summary, we have shown that, by employing entangled-photon
pairs that are generated by type-II SPDC in a non-collinear
configuration, one can obtain absolute ellipsometric data from a
reflective sample. The underlying physics that permits such
ellipsometric measurements resides in the fact that fourth-order
(coincidence) quantum interference of the photon pairs, in
conjunction with polarization entanglement, emulates an idealized
classical ellipsometric setup that utilizes a source and a
detector that are both calibrated absolutely.

This work was supported by the National Science Foundation and by
the Center for Subsurface Sensing and Imaging Systems (CenSSIS),
an NSF engineering research center. A. V. Sergienko's e-mail
address is alexserg@bu.edu.

%\newpage
%%
%% Figures and tables appear after References
%% One or more pages listing all figure captions should be
%% followed by the figures, one to a page.  Multipart figures
%% (2a, 2b, 2c, ...) may be placed on the same page.
%%
%% Following the figures, all tables should be placed
%% one table per page, except for long tables, which are allowed
%% to occupy multiple pages if necessary.  Table captions
%% should appear above the table to which it refers.
%%
%\section*{List of figures}

\vskip.5in
\begin{figure}[h]
\caption{Entangled-photon quantum ellipsometer. NLC stands for
nonlinear crystal; $A_1$ and $A_2$ are linear polarizers; $D_1$
and $D_2$ are single-photon detectors; and $N_c$ is the
coincidence rate. The sample is characterized by the ellipsometric
parameters $\psi$ and $\Delta$.  }
\end{figure}

\begin{figure}[h]
\caption{Unfolded version of the entangled-photon quantum
ellipsometer displayed in Fig. 1. }
\end{figure}
\newpage

\begin{figure}[h]
 \centering
 \epsfxsize=3.4in \epsfysize=2.6in
 \epsffile{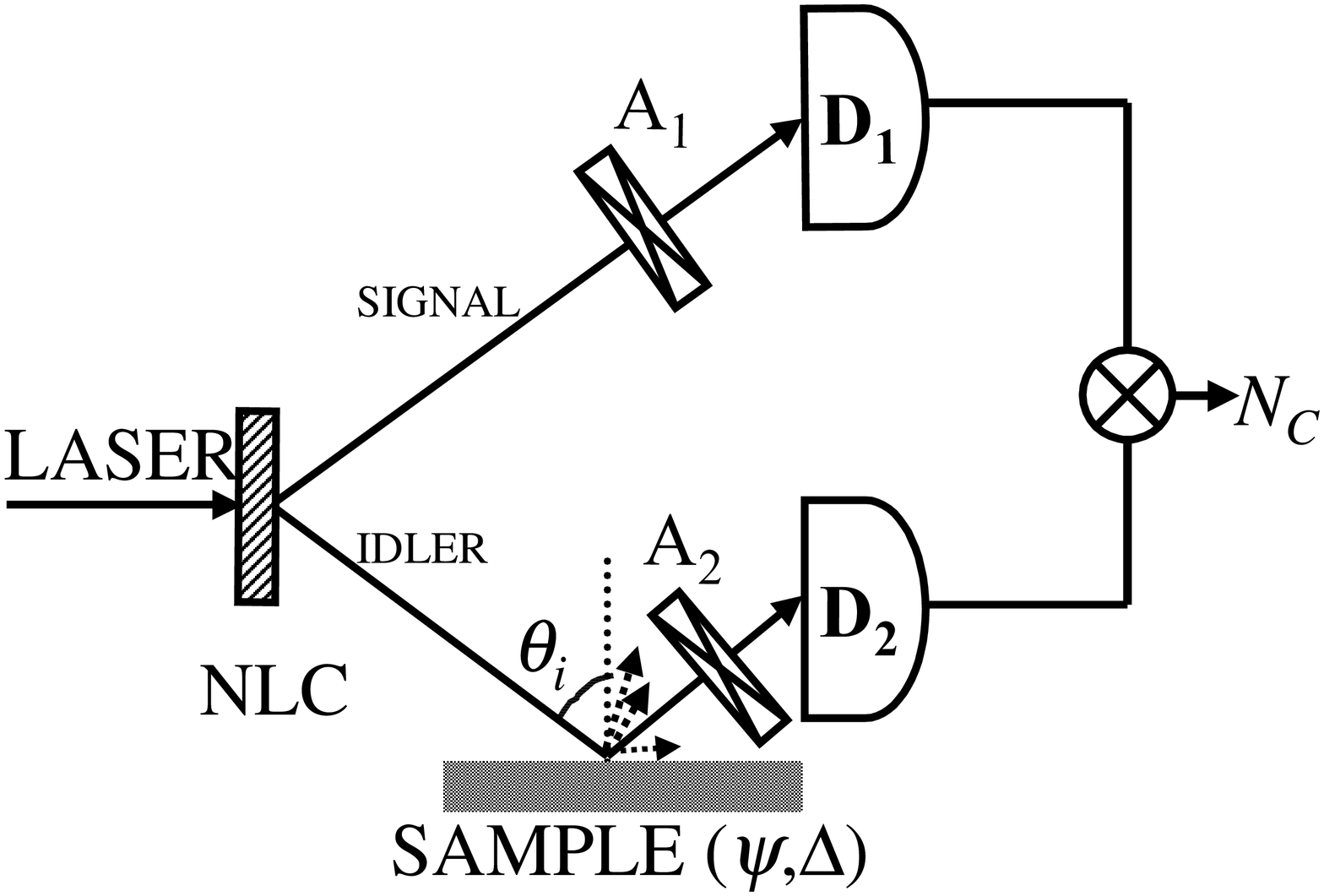}
 \vglue 0.2cm
 \label{Fig.1}
 \end{figure}

\vskip3in
Figure 1, A. F. Abouraddy et al. %% A short identifyer should appear
                                  %% at the bottom of the page with the
                                  %% figure, indicating figure number and
                                  %% first author or manuscript number
%\newpage

\vskip.5in
\begin{figure}[h]
 \centering
 \epsfxsize=3.4in \epsfysize=2.6in
 \epsffile{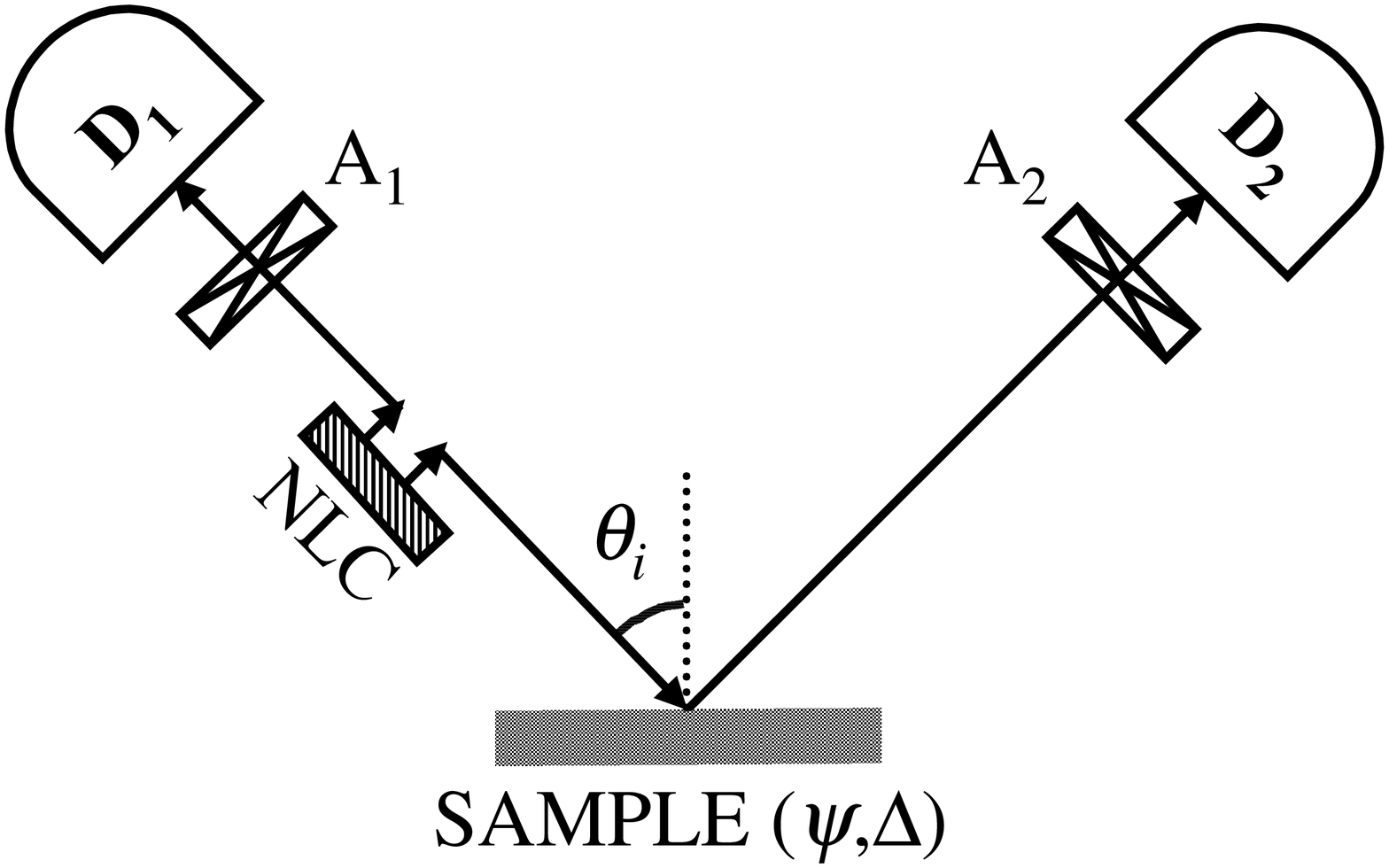}
 \vglue 0.2cm
 \label{Fig.2}
 \end{figure}

\vskip3in
Figure 2, A. F. Abouraddy et al. %% A short identifyer should appear
                                  %% at the bottom of the page with the
                                  %% figure, indicating figure number and
                                  %% first author or manuscript number

\end{document}